\documentclass[a4paper,fleqn,usenatbib]{mnras}

\usepackage[T1]{fontenc}
\usepackage{ae,aecompl}

\usepackage{graphicx}	
\usepackage{amsmath}	
\usepackage{amssymb}	
\usepackage{enumerate}
\usepackage{epsfig,graphicx,natbib,amsmath,amsfonts,amssymb}

\newcommand{\msun}{$\rm M_\odot$}
\newcommand{\mstar}{$M_*$}

\newcommand{\dindex}{D$_n$4000} 
\newcommand{\hd}{H$\delta$}                  
\newcommand{\hda}{\hd$_A$}
\newcommand{\ha}{H$\alpha$}
                    
\newcommand{\ewhae}{\rm{EQW$_{\rm H\alpha}$}}

\newcommand{\oii}{[{O~\sc ii}]$\lambda$3727}
\newcommand{\oiii}{[{O~\sc iii}]$\lambda\lambda$4959,5007}
\newcommand{\nii}{[{N~\sc ii}]$\lambda$6584}
\newcommand{\sii}{[{S~\sc ii}]$\lambda\lambda$6717,6731}

\title[MaNGA: Properties of kinematically decoupled galaxies]{SDSS-IV MaNGA: Properties of galaxies with kinematically decoupled stellar and gaseous components}

\author[Y. Jin et al.]{Yifei Jin$^{1,2,3}$, Yanmei Chen$^{1,2,3}$\thanks{E-mail: email: chenym@nju.edu.cn}, Yong Shi$^{1,2,3}$, C. A. Tremonti$^{4}$, M. A. Bershady$^{4}$,
\newauthor 
M. Merrifield$^{5}$, E. Emsellem$^{6,7}$, Hai Fu$^{8}$, D. Wake$^{4,9}$, K. Bundy$^{10}$, Lihwai Lin$^{11}$, 
\newauthor
M. Argudo-Fernandez$^{12,18}$, Song Huang$^{10}$, D. V. Stark$^{10}$, T. Storchi-Bergmann$^{13,22}$, 
\newauthor
D. Bizyaev$^{14,15}$, J. Brownstein$^{16}$, J. Chisholm$^{4}$, Qi Guo$^{17}$, Lei Hao$^{18}$, Jian Hu$^{19}$, 
\newauthor
Cheng Li$^{19}$, Ran Li$^{17}$, K. L. Masters$^{20}$, E. Malanushenko$^{14}$, Kaike Pan$^{14}$,  
\newauthor 
R. A. Riffel$^{21,22}$, A. Roman-Lopes$^{23}$, A. Simmons$^{14}$, D. Thomas$^{20}$, Lan Wang$^{17}$, 
\newauthor
K. Westfall$^{21}$, and Renbin Yan$^{24}$
\\ \\
$^{1}$School of Astronomy and Space Science, Nanjing University, Nanjing 210093, China\\
$^{2}$Key Laboratory of Modern Astronomy and Astrophysics (Nanjing University), Ministry of Education, Nanjing 210093, China\\
$^{3}$Collaborative Innovation Center of Modern Astronomy and Space Exploration, Nanjing 210093, China\\
$^{4}$Astronomy Department, University of Wisconsin, 475 N. Charter St., Madison, WI 53711, USA\\
$^{5}$School of Physics and Astronomy, University of Nottingham, University Park, Nottingham NG7 2RD, UK\\
$^{6}$European Southern Observatory, Karl-Schwarzschild-Str. 2, D-85748 Garching, Germany\\
$^{7}$Universit\'{e} Lyon 1, Observatoire de Lyon, Centre de Recherche Astro- physique de Lyon and Ecole Normale Sup\'{e}rieure de Lyon,\\
          \hspace{0.4cm}9 avenue Charles Andr\'{e}, F-69230 Saint-Genis Laval, France\\
$^{8}$Department of Physics \& Astronomy, University of Iowa, 751 Van Allen Hall, Iowa City, IA 52242, USA\\
$^{9}$Department of Physical Sciences, The Open University, Milton Keynes, MK7 6AA, UK\\
$^{10}$Kavli Institute for the Physics and Mathematics of the Universe, Todai Institutes for Advanced Study, the University of Tokyo,\\
          \hspace{0.4cm}277-8583 Kashiwa, Japan, (Kavli IPMU, WPI)\\
$^{11}$Academia Sinica Institute of Astronomy and Astrophysics, P.O. Box 23-141, Taipei 10617, Taiwan\\
$^{12}$Universidad de Antofagasta, Unidad de Astronom\'{i}a, Facultad Cs. B\'{a}sicas, Av. U. de Antofagasta 02800, Antofagasta, Chile\\
$^{13}$Departamento de Astronomia, Universidade Federal do Rio Grande do Sul, IF, CP 15051, 91501-970 Porto Alegre, RS, Brazil\\
$^{14}$Apache Point Observatory and New Mexico State University, PO Box 59, Sunspot, NM 88349-0059, USA\\
$^{15}$Sternberg Astronomical Institute, Moscow State University, Moscow, Russia\\
$^{16}$Department of Physics and Astronomy, University of Utah, 115 S 1400 E, Salt Lake City, UT 84112, USA\\
$^{17}$National Astronomical Observatories, Chinese Academy of Sciences, 20A Datun Road, Chaoyang, Beijing 10012, China\\
$^{18}$Shanghai Astronomical Observatory, Chinese Academy of Science, 80 Nandan Road, Shanghai 200030, China\\
$^{19}$Tsinghua Center of Astrophysics \& Department of Physics, Tsinghua University, Beijing 100084, China\\
$^{20}$Institute of Cosmology and Gravitation, University of Portsmouth, Portsmouth PO1 3FX, UK\\
$^{21}$Departamento de F\'{i}sica, Centro de Ci\^{e}ncias Naturais e Exatas, Universidade Federal de Santa Maria, 97105-900, Santa Maria, RS, Brazil\\
$^{22}$Laborat\'{o}rio Interinstitucional de e-Astronomia - LIneA, Rua Gal. Jos e Cristino 77, Rio de Janeiro, RJ - 20921-400, Brazil\\
$^{23}$Departamento de F\'{i}sica, Facultad de Ciencias, Universidad de La Serena, Cisternas 1200, La Serena, Chile\\
$^{24}$Department of Physics and Astronomy, University of Kentucky, 505 Rose St., Lexington, KY 40506-0057, USA
}
\date{Accepted XXX. Received YYY; in original form ZZZ}

\pubyear{2016}

\begin{document}
\label{firstpage}
\pagerange{\pageref{firstpage}--\pageref{lastpage}}
\maketitle

\begin{abstract}
We study the properties of 66 galaxies with kinematically misaligned gas and stars 
from MaNGA survey. The fraction of kinematically misaligned galaxies varies with 
galaxy physical parameters, i.e. \mstar, SFR and sSFR. According to their 
sSFR, we further classify these 66 galaxies into three categories, 
10 star-forming, 26 ``Green Valley'' and 30 quiescent ones. The properties of different types 
of kinematically misaligned galaxies are different in that the star-forming ones have 
positive gradient in \dindex\ and higher gas-phase metallicity, while the green valley/quiescent ones 
have negative \dindex\ gradients and lower gas-phase metallicity on average. There is evidence that all types of the kinematically 
misaligned galaxies tend to live in more isolated environment. Based on all these observational results, we propose a scenario 
for the formation of star forming galaxies with kinematically misaligned gas and stars $\--$ the progenitor accretes 
misaligned gas from a gas-rich dwarf or cosmic web, the cancellation of angular momentum from gas-gas collisions 
between the pre-existing gas and the accreted gas largely accelerates gas inflow, leading to fast centrally-concentrated 
star-formation. The higher metallicity is due to enrichment from this star formation.
For the kinematically misaligned green valley and quiescent galaxies, they might be formed through 
 gas-poor progenitors accreting kinematically misaligned gas from satellites which are smaller in mass.

\end{abstract}


\section{Introduction}
 \label{sec:intro}
 
In the framework  of hierarchical structure formation,  a galaxy grows
from primordial  density fluctuations and its  subsequent evolution is
shaped by  a range of  internal and external processes.   Galaxies can
recycle  gas  through stellar  mass  loss.   Stellar evolution  models
predict that a stellar population on average returns about half of its
stellar  mass   to  the   interstellar  space   over  a   Hubble  time
\citep{Jungwiert01,  Lia02, Pozzetti07}.   As a  result of  the angular
momentum conservation,  such gas should be  kinematically aligned with
the stars.  On the other hand, external processes, e.g. major mergers,
minor mergers  and gas  accretion from  the halo,  could also  bring a
significant amount of  gas into a galaxy to reshape  its structure and
evolution.  If externally acquired gas  can enter a galaxy with 
random angular  momentum, a mismatch in  the  star and  gas kinematics  is
expected \citep{bertola-1992a, bureau-2006, corsini-2014}. 
Galaxies  with misaligned gas and  stellar kinematics  are  the key
demonstrations    for   the    regulation   by    external   processes
\citep{rubin1994, Schweizer98}.

The phenomenon  of gas and  star misalignment  is now known  to be
ubiquitous  in elliptical  and lenticular  galaxies. Systematic    studies   with   long-slit
spectroscopy   have   reported   a    fraction   of $\sim$20--25 per cent
\citep{bertola-1992a, Kuijken96, kannappan-2001}. This fraction increases to a value
of $\sim$30-40 per cent with integral-field  spectroscopy  \citep{Sarzi06,  davis-2011,  barrera-2014,  barrera-2015}. 
However, the frequency of this phenomenon in the general galaxy 
population is still uncertain. There are only two statistical estimates 
of the frequency of counter rotators in a general population, and the sample 
size is about tens of galaxies \citep{kannappan-2001, pizzella-2004}. Both works 
report zero detection of counter rotators in star forming galaxies. Equally important, 
we have no idea about whether the physical mechanisms responsible for misaligned gas-star kinematics
completely reshape the host galaxies or merely perturb them; whether there is any 
difference in the physical properties between blue and red misaligned galaxies, are they 
on an evolution sequence, or related in some way?

In this paper, we search for galaxies with decoupled gas and star kinematics from MaNGA 
\citep[Mapping Nearby Galaxies at Apache Point Observatory,][]{bundy-15}, a new integral field spectroscopic survey. 
The sample size of MaNGA is about ten times larger than that of ATLAS$^{\rm3D}$ \citep{cappellari-2011} 
and CALIFA \citep{sanchez-2012}. 
In contrast to long-slit spectroscopy which fails to identify the pattern of the kinematic misalignment between gas and stars out of the complicated kinematics in Spirals especially barred Spirals, the IFU observation is important in obtaining more robust determinations of
misaligned gas/star kinematics.
Finally, the MaNGA sample is unbiased with respect to
morphology, inclination, color, etc. And it can be trivially corrected to a volume complete sample.
This is the first time we study this kind of galaxies in a large and complete sample with IFU observation.
\cite{chen-prep} report the fraction of kinematic misalignment phenomenon 
in different categories of galaxies: star-forming sequence, green valley and quiescent sequence.  
They focus on the misaligned star forming galaxies, finding their central regions undertake more intense star-formation than the outskirt, 
indicating the acquisition of external gas might be an important mechanism in the growth of the central regions.
As a follow-up work, we look into the properties of misaligned galaxies in all these three categories. 
This paper is organized as follows. 
In Section 2, we briefly introduce
the MaNGA survey and our data analysis method as well as sample selection criteria. 
The properties of the kinematically misaligned galaxies, including the fraction of misaligned galaxies 
as a function of galaxies physical parameters (\mstar, SFR, sSFR), the stellar population, metallicity as well as 
environment, are studied in Section 3. We discuss the observational results in Section 4. 
A short summary is presented in Section 5.
We use the cosmological parameters $H_{\rm0}$ = 70 km s$^{-1}$ Mpc$^{-1}$,
$\Omega_{\rm M}$ = 0.3, and $\Omega_{\rm \Lambda}$ = 0.7 throughout this paper.

\section{Data}
\subsection{The MaNGA Survey}
MaNGA is one of three core programs in the fourth-generation Sloan Digital Sky Survey (SDSS-IV) 
that began on 2014 July 1 \citep{bundy-15, drory-2015, yan-2016}, using SDSS 2.5-in telescope \citep{gunn-2006}. It is designed to investigate the internal kinematic structure and composition 
of gas and stars in an unprecedented sample of $\sim$10,000 nearby galaxies at a spatial resolution of 2$^{\prime\prime}$ ($\sim$ 1kpc). MaNGA employs dithered 
observations with 17 fiber-bundle integral field units with 5 sizes: 2$\times$N$_{19}$ (12$^{\prime\prime}$ in diameter),
4$\times$N$_{37}$,4$\times$N$_{61}$,2$\times$N$_{91}$, 5$\times$N$_{127}$ (32$^{\prime\prime}$). 
Two dual-channel BOSS spectrographs \citep{smee-2013} provide simultaneous wavelength coverage 
over 3600$\sim$10300$\rm \AA$ at R $\sim$ 2000. 
With a typical integration time of 3 hours, MaNGA reaches a target 
$r-$band signal-to-noise ratio of 4$\sim$8 ($\rm \AA$$^{-1}$ per fiber) at 23 ABmag arcsec$^{-2}$, which is typical for 
the outskirts of MaNGA galaxies.  The observing strategy for MaNGA can be found in \cite{Law-15}.

The MaNGA targets constitute a luminosity-dependent volume-limited sample, selected 
almost entirely from SDSS ``main" galaxy sample \citep{strauss-2002}. An intensive effort to design and optimize the 
MaNGA target selection under various constraints and requirements is presented in D. A. Wake et al. 
(in preparation), who also give details on the final selection cuts applied and the resulting sample sizes 
and their distributions in various properties. In what follows, we present a brief conceptual summary 
based on a near-final set of selection criteria. The MaNGA sample design is predicated on three major 
concepts: (1) the sample size is required to be roughly 10,000 galaxies; (2) the sample is complete 
above a given stellar mass limit of log\mstar/\msun > 9, and has a roughly flat log$M_*$ distribution. 
This selection is motivated by ensuring adequate sampling of stellar mass, which is widely recognized as a key parameter or 
``principal component" that defines the galaxy population; (3) the sample is chosen to prioritize uniform radial coverage in 
terms of the scale length associated with the galaxy's light profile, which is the major-axis 
half-light radius, $R_{\rm e}$. 

MaNGA selects ``Primary" and ``Secondary" samples defined by two radial coverage goals. The Primary selection 
($<$z$>$= 0.03) reaches 1.5 $R_{\rm e}$ (for more than 80 per cent of its targets) sampled with an average of five radial bins. 
It accounts for $\sim$5000 galaxies. To the Primary selection we add a ``Color-Enhanced sample" (CE) of an additional 
$\sim$1700 galaxies designed to balance the color distribution at fixed $M_*$. The color enhancement increases 
the number of high-mass blue galaxies, low-mass red galaxies, and ``green valley" galaxies tracing important but 
rare phases of galaxy evolution. We refer to these two selections as ``Primary+". The Secondary selection of $\sim$3300 
galaxies ($<$z$>$ = 0.045) is defined in an identical way to the Primary sample, but with a requirement that 80 per cent of 
the galaxies be covered to 2.5 $R_{\rm e}$. In this work, we use 1351 galaxies that have been observed by MaNGA till the summer
of 2015. The 1351 galaxies here constitute the first year of MaNGA observing. MaNGA targets are observed plate by plate, so and subset selected this way will have  the exact same distribution of IFU sizes as the final sample and should be a representative subset of the final sample.

\subsection{Data Analysis}
\subsubsection{Spectral fitting}

Principal component analysis (PCA) is a standard multivariate
analysis technique, designed to identify correlations in large data
sets. Using PCA, \cite{chen-2012} designed a method to estimate stellar masses, 
mean stellar ages, star formation histories (SFHs), dust extinctions and stellar velocity dispersions 
for galaxies from the Baryon Oscillation Spectroscopic Survey (BOSS). 
To obtain these results, we use the stellar population synthesis models 
of \cite[BC03]{bc03} to generate a library of model spectra with a broad range of SFHs,
metallicities, dust extinctions and stellar velocity dispersions. The PCA is run on this library
to identify its principal components (PC) over a certain rest-frame wavelength range 3700$-$5500$\rm \AA$. 
We then project both the model spectra and the observed spectra onto the first seven PCs to 
get the coefficients of the PCs, which represents the strength of each PC presented in the model or observed 
spectra. We derive statistical estimates of various physical parameters
by comparing the projection coefficients of the observed galaxy to those of the models as follows.
The $\chi^2$ goodness  of  fit  of  each  model  determines  the
weight $\sim {\rm exp(-\chi^2/2)}$ to be assigned to the physical parameters of that model when building the probability distributions
of the parameters of the given galaxy. The probability density
function (PDF) of a given physical parameter (in this work it is stellar velocity dispersion) is thus obtained
from the distribution of the weights of all models in the library.
We characterize the PDF using the median and the 16 per cent $-$84 per cent
range  (equivalent  to $\pm$1$\sigma$ range  for  Gaussian distributions).

In this work, we directly use the PCs given by \cite{chen-2012}. 
In order to get the velocity field of a galaxy, we shift the best fit model from $-$1000~km~s$^{-1}$ to 1000~km~s$^{-1}$ by a step size 
of 2~km~s$^{-1}$. For each step, we calculate reduced $\chi^2$ between the best fit model and the observed spectrum, the rotation velocity 
of stars $V_{\rm rot}^{\rm star}$ at a certain pixel is determined by the fit with the lowest $\chi^2$ value. The uncertainty of the 
rotation velocity is given by the width of the $\chi^2$ minimum at which $\Delta\chi^2$ = 1. Comparing with other codes \citep[e.g. pPXF,][]{ppxf} 
designed for extracting the stellar kinematics and stellar population from  absorption-line spectra of galaxies, 
the advantage of PCA is that it includes PCs which carry the 
information of stellar velocity dispersion, 
and the small and non-degenerate template set (PCs) makes it possible to obtain good
velocity dispersion measurements on low S/N data, removing the need for spatial binning of the data.
On the other hand, since we do not need to fit stellar velocity dispersion as a free parameter
like $V_{\rm rot}^{\rm star}$, the PCA method transfers a two-dimensional numerical problem into one-dimension, largely 
decreasing the calculation time. The disadvantage is that it can not give h3 and h4 Gauss-Hermite moments.

\begin{figure*}
  \begin{center}
    \epsfig{figure=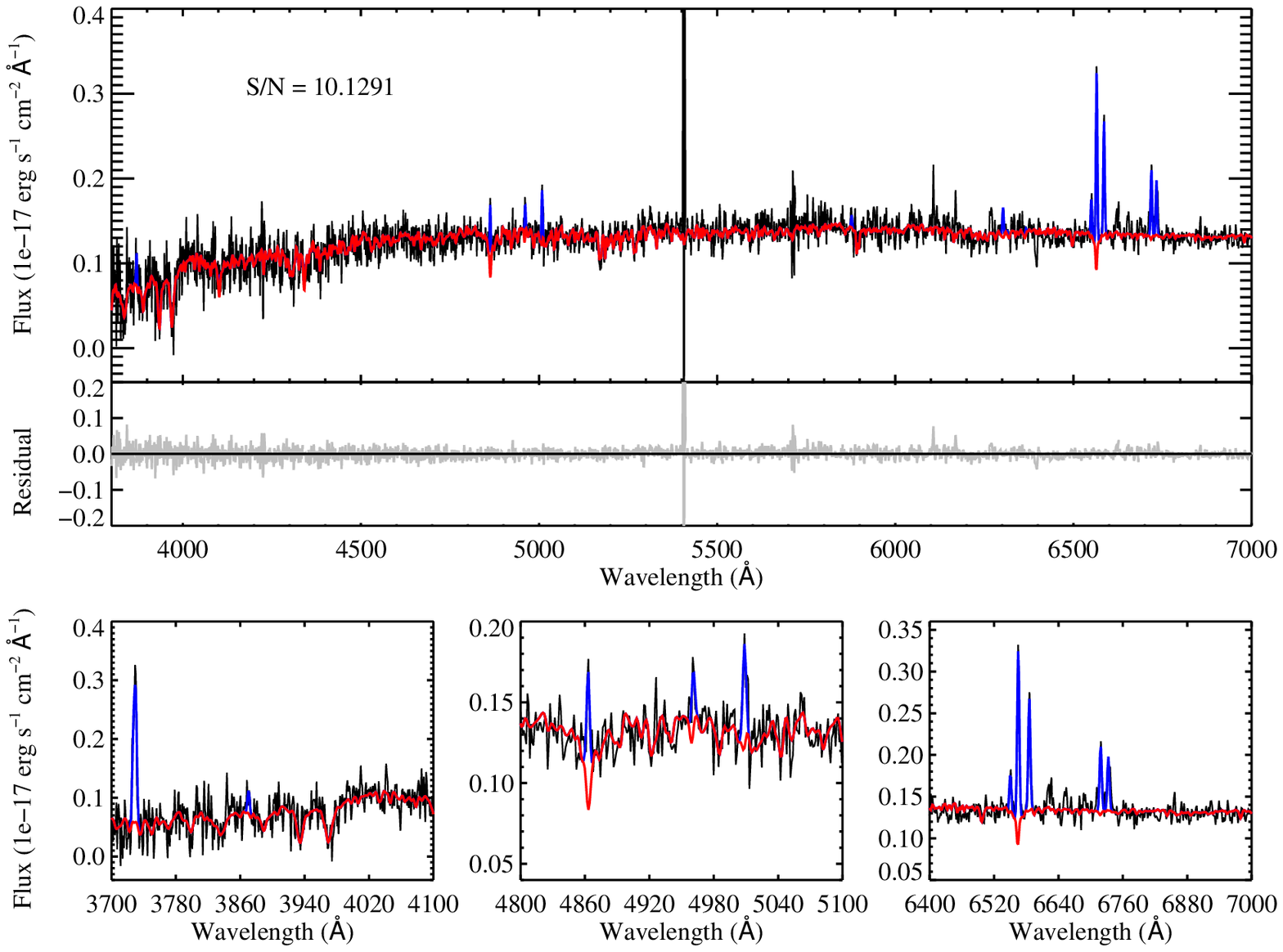,clip=true,width=0.8\textwidth}  
    
  \end{center}
  \caption{One example of spectral fitting. The top panel shows the observed spectrum (black), best-fitting model continuum (red) and the single Gaussian fitted emission lines (blue). Residuals are presented in the middle plane. The bottom three windows zoom in \oii, \oiii\ and H$\alpha$ emission line regions, showing the details of the fit.}
  \label{fig:spectra_fit}
\end{figure*}

Once we have the rotation velocity and stellar velocity dispersion for each spatial pixel,
we  model  the  stellar  continuum  of  each spectrum using the BC03 stellar population synthesis models 
to separate  the  stellar continuum and absorption lines from the nebular emission lines.
The best fit continuum model is then subtracted from the observed spectrum 
to give the pure emission line spectrum, and each 
emission line is then fitted with one Gaussian component. 
We include \oii, H$\beta$, \oiii, H$\alpha$, \nii, \sii\ in our emission line fitting.
The line center and width of them are all tied together.
Figure~\ref{fig:spectra_fit} shows one example of the continuum and emission line fit. The black line is the observed spectrum with a median S/N $\sim$10 and the best fit model is in red. The blue components are the best Gaussian fit of emission lines. 
We derive the velocity field of ionized gas by comparing the line centers of emission lines to the rest-frame values. 

\subsubsection{Velocity field}\label{sec:rotmap}
\begin{figure*}
  \begin{center}
    \epsfig{figure=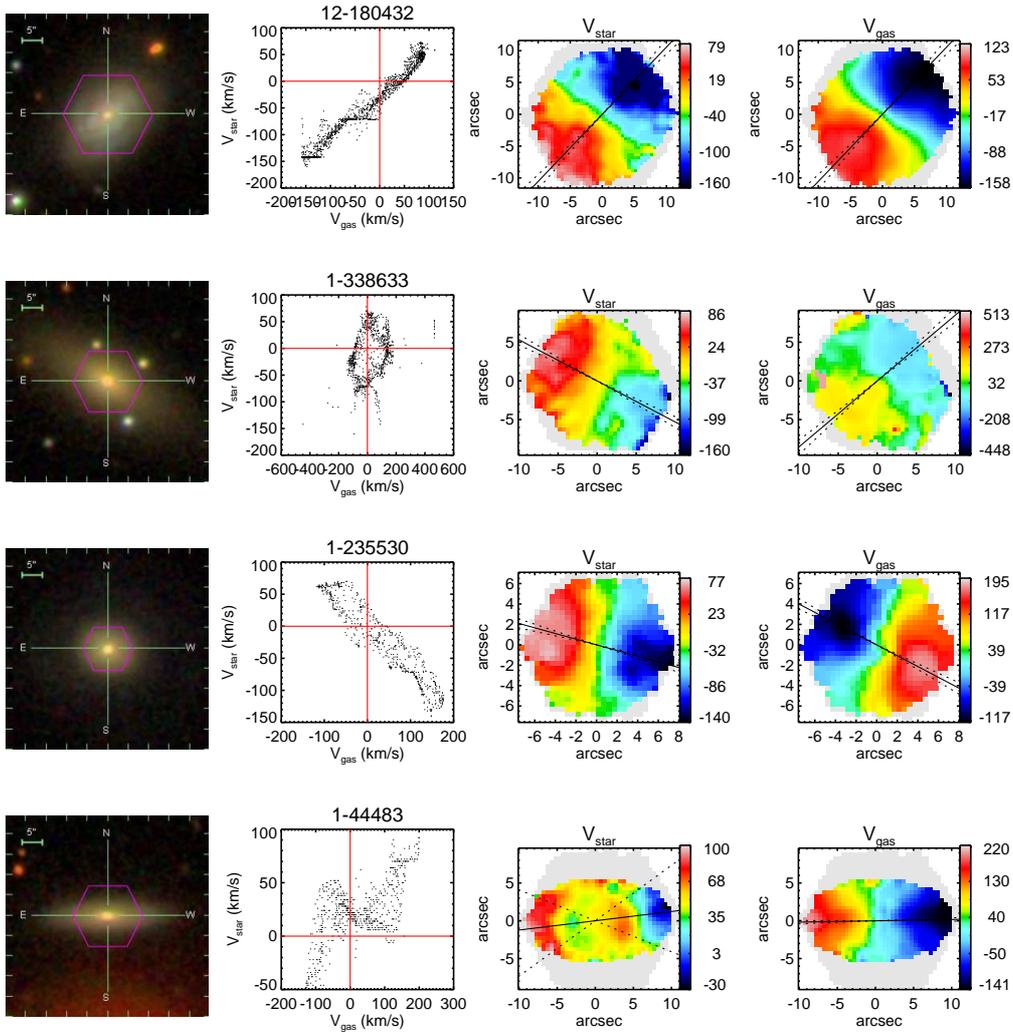,clip=true,width=0.8\textwidth}  
    
  \end{center}
  \caption{Four examples of velocity fields. Each row represents one type of galaxies as mentioned in Section~\ref{sec:rotmap}. The left column shows the SDSS false-color image in which the purple hexagon marks the region covered by MaNGA bundle. The middle column is the $V_{\rm star}$ versus $V_{\rm gas}$ diagram where the gaseous and stellar velocities are measured in the same spaxel. We use the velocity of H$\alpha$ to trace the motion of ionized gas. The red lines mark the zero values of the velocity for each component. 
The third and fourth columns show the stellar and gas velocity fields, respectively. We only show spaxels with spectral median S/N $>$ 5 in this plot. The color code indicates the values of velocity. The red side is moving away from us and the blue side is approaching. The solid line over-plotted in each velocity field shows the major axis of kinematic position angle fitted by FIT\_KINEMATIC\_PA \citep{krajnovic-2006}, while two dashed lines show $\pm1\sigma$ error range.}
  \label{fig:example_galaxy}
\end{figure*}

In Figure~\ref{fig:example_galaxy}, we show the velocity fields of stars and ionized gas of four galaxies as examples.
The top panel shows a galaxy in which the kinematics of gaseous and stellar 
components are aligned with each other; in the second row, we show a galaxy with gas and stars  
rotating perpendicularly; in the third row, it is an example of counter rotating 
galaxy; while the fourth row shows a 2$\sigma$ galaxy as that was found from ATLAS$^{\rm 3D}$ 
\citep{krajnovic-2011}. 2$\sigma$ galaxies are galaxies which are the superposition 
of two counter-rotating stellar disks. This produces a special velocity field with 
counter-rotation stellar core. The stellar velocity dispersion maps usually have two peaks along the 
major-axis away from the center, at the position where the two disks 
have the maximum difference in rotation velocity (which increases the value of stellar velocity dispersion), 
hence the name ``2$\sigma$". These galaxies are formed either by accretion 
of counter-rotating gas, or by the merging of two coplanar disks. 
The direction of the angular momentum vector of any tracer 
can be found by estimating its kinematic position angle (PA). 
The kinematics PAs of warm ionized gas and stars are measured using the FIT\_KINEMATIC\_PA 
routine described in appendix C of \cite{krajnovic-2006}. 
It is defined as the counter-clockwise angle between north and a line which bisects 
the tracer's velocity field, measured on the receding side. One can then define the 
kinematic misalignment angle ${\rm \Delta PA}$ = $|$${\rm PA}_{\rm star} -  {\rm PA}_{\rm gas}$$|$ as the difference between 
the kinematic PAs of gas and stars.
The solid lines in Figure 2 show the best fit position 
angles and the two dashed lines show the $\pm 1\sigma$ error. We define galaxies with 
${\rm \Delta PA}>30^\circ$ as gas-star misaligned ones. 

\subsubsection{Star formation rates $\&$ stellar masses}\label{sec:sfrsm}

\cite{chang-2015} produced new calibrations for global SFR and \mstar\ by combining SDSS and WISE photometry for the SDSS spectroscopic galaxy sample. They created the spectral energy distributions (SEDs) that cover $\lambda$ = 0.4-22 $\mu m$ for 858,365 present-epoch galaxies. Using MAGPHYS, they then modeled the attenuated stellar SED and dust emissions at 12 and 22 $\mu m$, calibrating the global SFR as well as \mstar.
We match the MaNGA galaxies with Chang's catalog to get the global SFR and stellar mass, finding 1220 matches. 
According to our definition, 66 of them are star-gas misaligned ones \citep{chen-prep}. 

\begin{figure*}
  \begin{center}
    \epsfig{figure=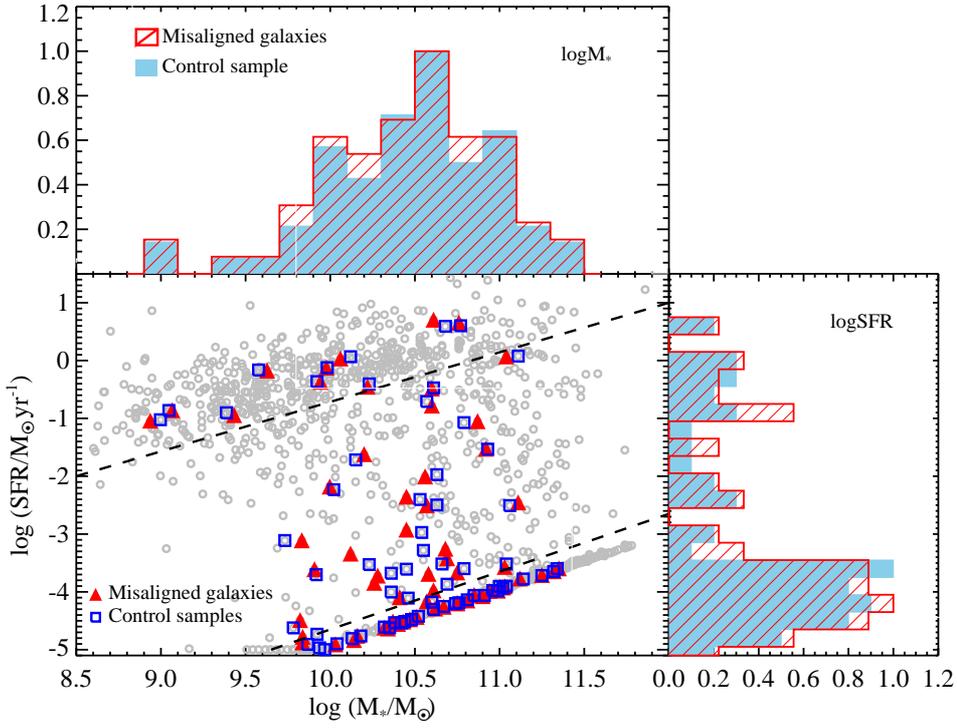,clip=true,width=0.8\textwidth}  
    
  \end{center}
  \caption{The SFR vs. \mstar\ diagram of 1220 MaNGA galaxies. The grey open circles represent 1220 MaNGA galaxies. 66 kinematically misaligned galaxies are presented as red filled triangles and the relevant control galaxies are shown as blue open squares. Two dashed lines are used to separate star-forming, green valley and quiescent galaxies \citep{chen-prep}. The histograms at top and right panels are the distributions of stellar masses and SFR, respectively. In each panel, the histogram filled with red lines is for the kinematically misaligned sample and that filled with blue color is for the control sample. The peaks of all the distributions are set to 1.}
  \label{fig:control_sample}
\end{figure*}

To investigate how the properties of misaligned galaxies are different from others, we built a control sample of 66 kinematically aligned galaxies as comparison.
For each misaligned galaxy, we select a kinematically aligned (${\rm \Delta PA}<30^\circ$) control galaxy which is closely matched in SFR and \mstar. The motivations for choosing this set of matching parameters are the following: 
(i) constraining the control galaxies to have similar stellar mass is extremely important because stellar population properties are known to vary strongly as a function of galaxy mass. 
(ii) constraining the control galaxies to have similar SFR is due to the fact that the fraction of kinematically misaligned galaxies is a strong function of star formation activity, see section~\ref{sec:fraction}.
Figure~\ref{fig:control_sample} shows the distributions of kinematically misaligned galaxies (red) and the control sample (blue) in SFR vs. \mstar\ diagram. 
The top panel gives the histograms of \mstar\ of misaligned sample (red) and control sample (blue). The right panel shows the histograms of SFR. The peaks of all the histograms are set to 1. The control galaxies we select match well with the misaligned galaxies in these two parameters.

We further classify 66 misaligned/control galaxies into 10 star-forming, 26 ``Green Valley'' and 30 quiescent galaxies based on their locations in SFR vs. \mstar\ diagram.
The star-forming galaxies are defined as $\rm{log SFR > 0.86 \times log M_* - 9.29}$, 
the quiescent ones have $\rm{log SFR < log M_* -14.65}$ and the galaxies in between are defined as green valley galaxies.
The two dashed lines in Figure~\ref{fig:control_sample} are to discriminate star-forming, green valley and quiescent galaxies.
These three categories of misaligned/control galaxies are used in following sections.


\section{Results}\label{sec:result}

\subsection {The fraction of kinematically misaligned galaxies as a function of galaxy physical parameters}\label{sec:fraction}


Thanks to MaNGA target selection strategy, it provides us with a representative sample of all galaxy types where the selection is 
well understood.
Also due to the large sample of MaNGA, for the first time, we can study
how the fraction of kinematic misalignment depends on galaxies physical parameters, i.e. \mstar, SFR, sSFR $\equiv$ SFR/\mstar, over all galaxy types.

\begin{figure*}
  \begin{center}
    \epsfig{figure=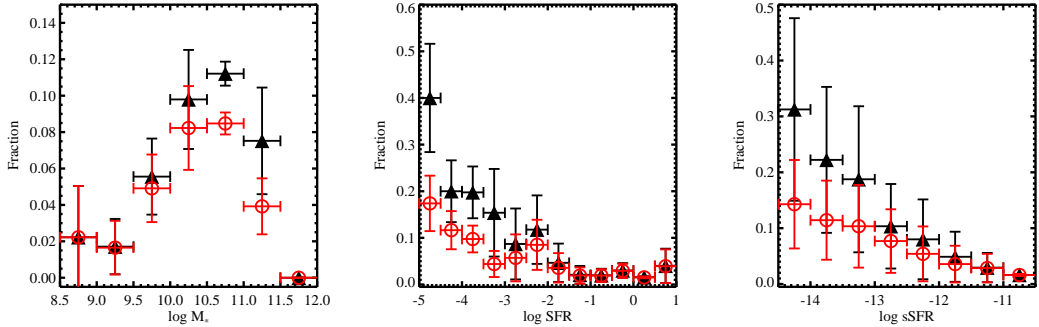,clip=true,width=.8\textwidth}  
    
  \end{center}
   \caption{The fraction of kinematically misaligned galaxies as a function of \mstar\ (left), SFR (middle) and sSFR (right). The black triangle is defined as $N(\rm \Delta PA > 30^\circ)$/$N$(EML) and the red open circle is defined as $N(\rm \Delta PA > 30^\circ)$/$N$(ALL). $N(\rm \Delta PA > 30^\circ)$ is number of kinematically decoupled galaxies in each parameter bin. $N$(EML) is number of galaxies with nebular emission lines. $N$(ALL) is total number of galaxies in a certain parameter bin (including galaxies with and without nebular emission). The x-axis error indicates the size of each parameter bin and y-axis error is estimated from bootstrap method.}
  \label{fig:fraction}
\end{figure*}

Figure~\ref{fig:fraction} shows the fraction of kinematically misaligned galaxies as a function of \mstar\ (left), SFR (middle) and sSFR (right). 
We estimate the sample variance errors using a bootstrap technique, by resampling the MaNGA galaxies for 2000 times. 
The errors are given by the root-mean-square difference between the fraction values calculated from each bootstrap sample and the mean value.
It is clear that this fraction peaks at log\mstar/\msun $\sim$ 10.5, and it is a decreasing
function of SFR as well as sSFR. In quality, the strong clustering
of the galaxies with decoupled gas and star kinematics in galaxies
with weak star formation activity can be easily understood -- if
all the galaxies accrete retrograde gas with equal probability, then
for the gas poor low SFR galaxies, the accreted gas would survive longer
since the interaction with existing gas is less important and the collision
cross section between gas and star is too small to influence the retrograde
angular momentum of the accreted gas. The reduction of this fraction at
low stellar mass end corresponds to the fact that in the local universe,
galaxies with log\mstar/\msun $<$ 10 are dominated by star forming ones with higher sSFR. 
The decreasing of this fraction at the high mass end is consistent with
the result of ATLAS$^{\rm 3D}$ \citep{davis-2011}, which is based on a sample of massive
early type galaxies (ETGs). They suggest that the alignment in the most massive
galaxies can be caused by galaxy scale processes which reduce the probability
that cold, kinematically misaligned gas can be accreted on to the galaxy, i.e.
AGN feedback, the ability to host a hot X-ray gas halo, or a halo mass threshold.

\subsection{Galaxy morphologies and H$\alpha$ EQW maps} \label{sec:morphology}

\begin{figure*}
  \begin{center}
    \epsfig{figure=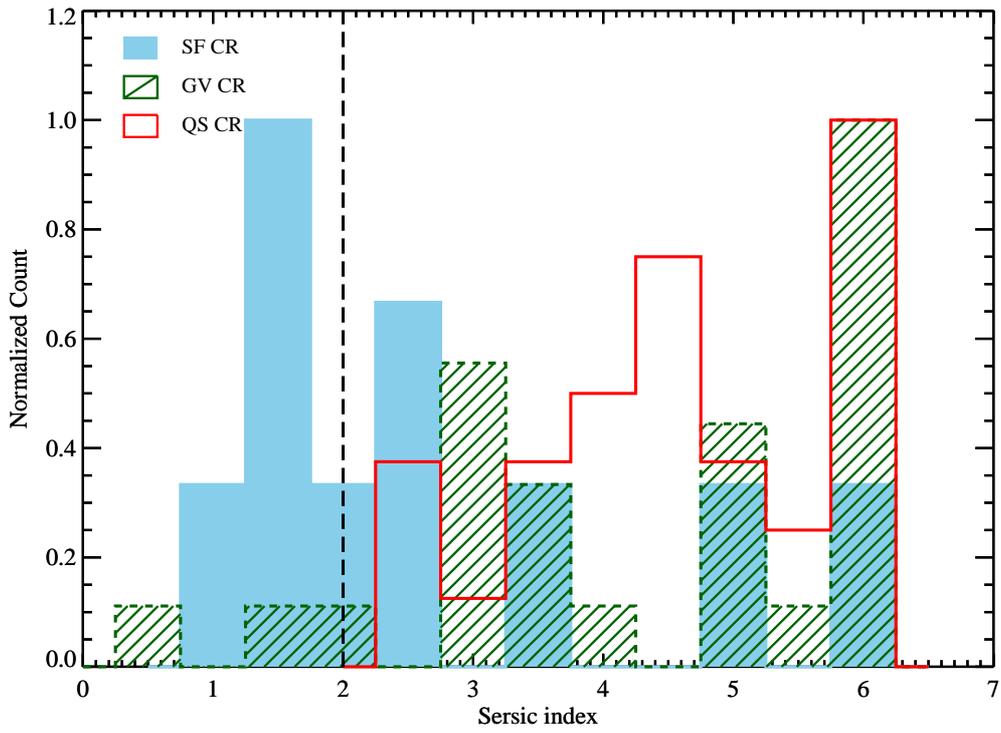,width=0.8\textwidth}  
    
  \end{center}
  \caption{The s$\acute{\rm e}$rsic index ($n$) distributions for star-forming (blue), green valley (green) and quiescent (red) misaligned galaxies. The peak of each distribution is set to 1. The vertical dashed line marks $n=2$, which is the often used proxy for bulge versus disk dominated galaxies.}
  \label{fig:sersic}
\end{figure*}

Before undertaking any further investigations on these kinematically misaligned galaxies, we look into their morphologies first. 
We use the s$\acute{\rm e}$rsic indices \emph{n} \citep{sersic-1963, sersic-1968} to characterize the morphologies of 66 kinematically misaligned galaxies. 
The surface brightness profile of galaxies can be fit by the S$\acute{\rm e}$rsic profile:
\begin{equation} \label{equ:sersic}
I(R) = I_{0}exp \left[-\beta_{n} {\left(\frac{R}{R_e}\right)}^{1/n}\right]
\end{equation}
where $I_{0}$ is the central surface brightness and $R_e$ is the effective radius that encloses half of the total light. 
The parameter $n$ is the well known s$\acute{\rm e}$rsic index. Figure~\ref{fig:sersic} shows the distribution of the 
s$\acute{\rm e}$rsic index taken from NASA-Sloan Atlas\citep{abazajian-2009,blanton-2011}\footnote{ http://www.nsatlas.org/} 
for star-forming (blue), green valley (green) and quiescent (red) kinematically misaligned galaxies.
The vertical dashed line marks the position of $n=2$, which is the often used proxy for bulge versus disk dominated galaxies. 
Misaligned star forming galaxies are indeed more disk like, while most green valley and quiescent misaligned galaxies seem to be spheroid dominated. However, this should be confirmed dynamically, e.g. via an objective assessment of the stellar angular momentum.

Sixty percent of the star-forming, 92\% green valley and all the quiescent misaligned galaxies have s$\acute{\rm e}$rsic indices $>$ 2. While the fraction of the overall galaxy population for each category is 27, 68 and 96\%, respectively. The s$\acute{\rm e}$rsic index distribution of star forming and green valley misaligned galaxies are significantly different from their general galaxy population.

\begin{figure*}
  \begin{center}
    \epsfig{figure=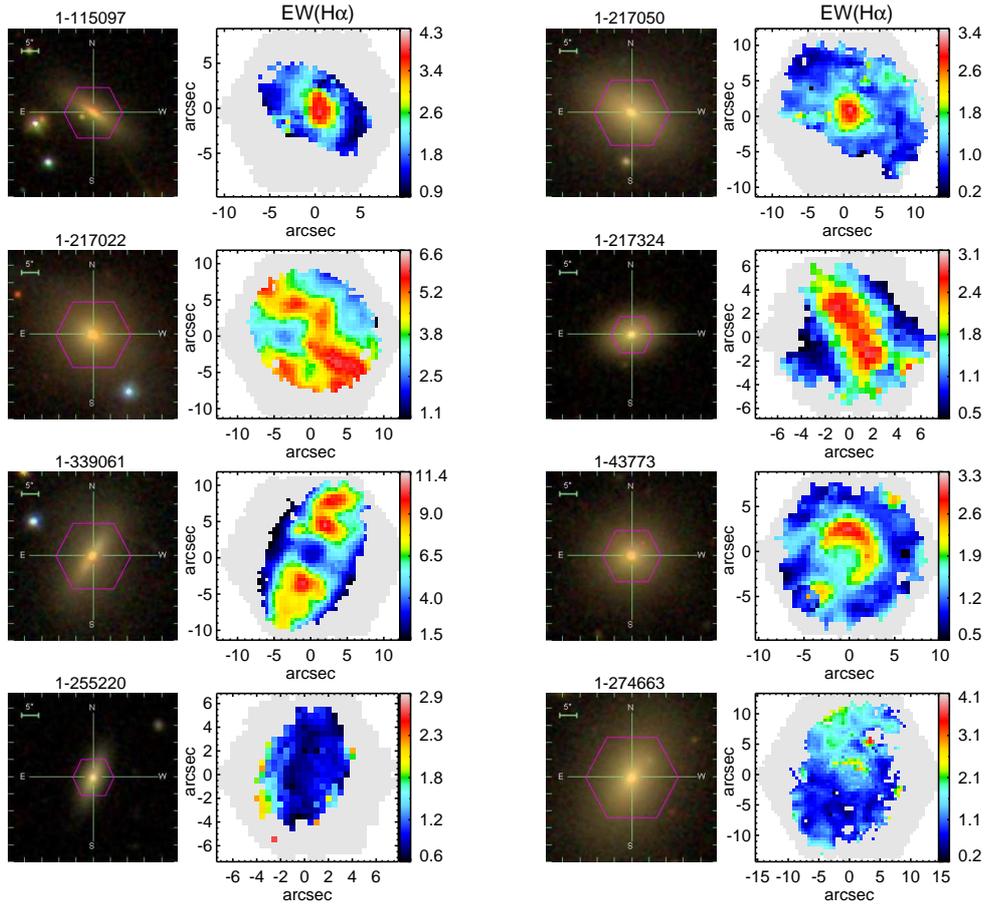,width=0.8\textwidth}  
    
  \end{center}
  \caption{The SDSS false-color images as well as the \ewhae\ maps of 8 kinematically misaligned galaxies. In the SDSS false-color images, the purple hexagon marks the region covered by MaNGA bundle. The color coded regions in \ewhae\ maps are spaxels with S/N $>$ 3 for \ha\ emission line. Background grey spaxels show the size of the bundle, corresponding to the regions marked by purple hexagons in the images. The galaxies in the same row belong to the same type of \ewhae\ morphological classification defined in Section~\ref{sec:morphology}.}
  \label{fig:morph}
\end{figure*}

In Figure~\ref{fig:morph}, we show the SDSS false-color images and the \ha\ equivalent width (\ewhae ) maps of 8 kinematically misaligned galaxies.
All the spaxels shown in the \ewhae\ maps have reliable emission line detections with S/N $>$ 3. 
We divide the 66 kinematically misaligned galaxies into the following four classes based on the substructures revealed in \ewhae\ maps. 

\begin{enumerate}[(1)]
\item About 23 per cent of misaligned galaxies have \ewhae\ peaking at center and decreasing with radius (top panel of Figure~\ref{fig:morph}).
\item \ewhae\ map has biconical patterns. These galaxies with old stellar populations are referred as the ``red geysers'' discovered by 
\cite{cheung-2016} in MaNGA survey, providing a strong indication on the material exchange processes such as gas accretion, minor merger and outflow (second panel). We only find 4 misaligned galaxies (6 per cent) with this feature.
\item There are 10/66 misaligned galaxies show ring-like or off-center structure in \ewhae\ map (third panel).
\item The whole \ewhae\ maps of 56 per cent misaligned galaxies are smooth, no obvious substructure exists (fourth panel).
\end{enumerate}

\subsection{Stellar population} \label{sec: sfh}
\subsubsection{\dindex\ \& \hda\ map}

We investigate the star formation history of kinematically misaligned galaxies 
using the \dindex\ versus \hda\ diagnostic diagram developed by 
\cite{kauffmann-2003b}. It is based on the measurement of two famous 
spectral indices around the 4000\AA\ wavelength region, the 4000\AA\ 
break (\dindex) and the strength of \hd\ absorption line (\hda). 
At wavelength around 4000\AA\ , galaxy spectra vary in both spectral 
shape and strength of \hd\ absorption line. These two are inversely 
correlated: younger galaxies have deeper \hd\ absorption and weaker 
4000\AA\ break. The combination of these two indices can tell us 
the recent star formation history of galaxies. 

We closely follow the measurements adopted by \cite{kauffmann-2003b}. 
The index \dindex\ is parameterized as the ratio of the flux density 
between two narrow continuum bands 3850 -- 3950 and 4000 -- 4100\AA. 
The \hda\ index is defined as the equivalent width of \hd\ absorption 
feature in the bandpass 4083 -- 4122\AA. We measure the parameters 
for each spaxel and place them in \dindex\ versus \hda\ diagram. 
Considering the uncertainty of the spectral fitting, we only use 
the spaxels with median signal-to-noise ratio per pixel greater than 10.

\begin{figure*}
  \begin{center}
    \epsfig{figure=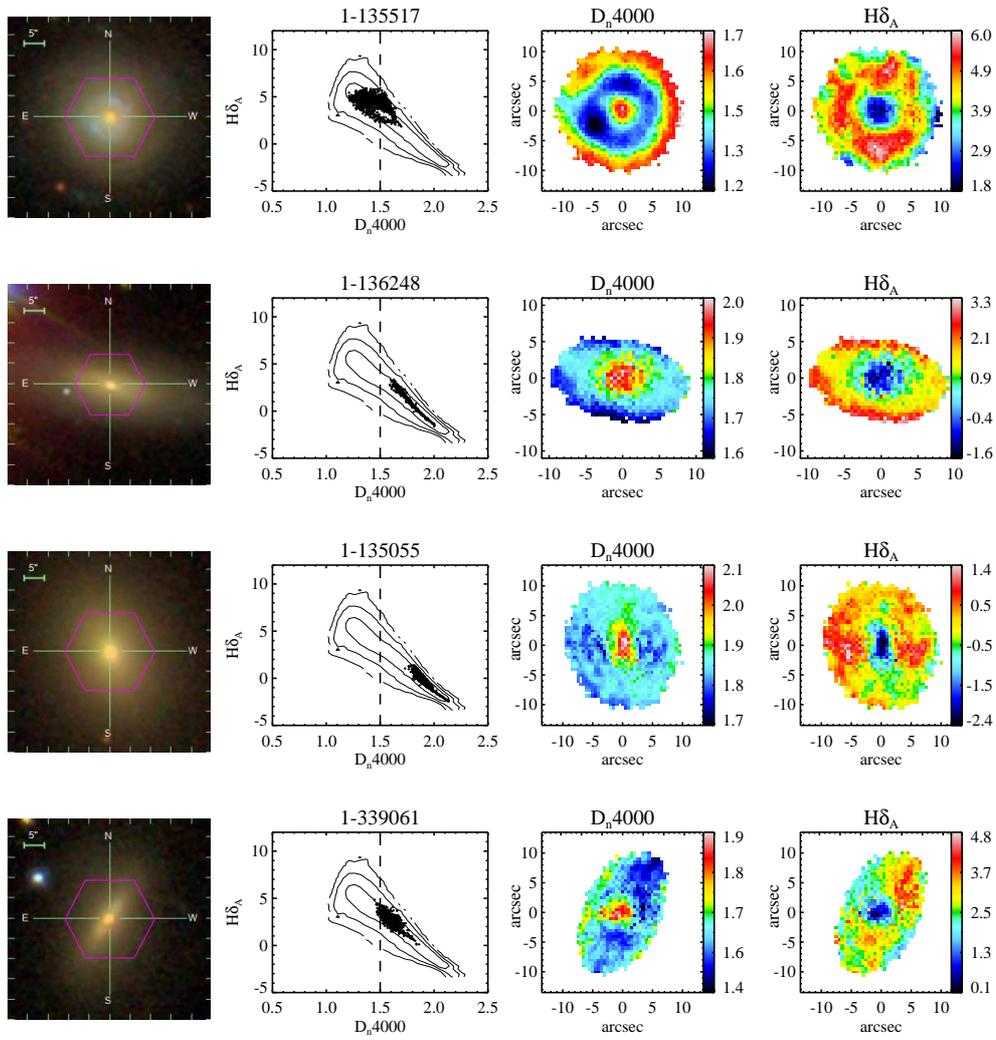,clip=true,width=0.8\textwidth}  
    
  \end{center}
  \caption{The spatial resolved \dindex\ and \hda\ distribution. The left column shows the SDSS false-color images. The second column shows the \dindex\ versus \hda\ diagrams. The background grey contours represent the distribution of the SDSS DR4 galaxies. The last two columns show the 2-D spatial resolved \dindex\ and \hda\ maps.}
  \label{fig:index}
\end{figure*}

Figure~\ref{fig:index} shows examples of spatial resolved \dindex\ and
\hda\ distribution. Different rows show representatives of different 
types of kinematically misaligned galaxies. From top to bottom, they are 
star-forming, green valley, quiescent galaxies, respectively. From left to right, the 
columns in Figure~\ref{fig:index} show the SDSS false-color images, 
the \dindex\ vs. \hda\ planes, as well as the \dindex\ and \hda\ maps, respectively.
The background contour is the distribution of SDSS DR4\footnote{http://wwwmpa.mpa-garching.mpg.de/SDSS/DR4/raw$\_$data.html} 
results in which the two parameters are measured in exactly the same way \citep{brinchmann-2004}. 
For green valley/quiescent galaxies, most of the spaxels have \dindex\ > 1.5, 
indicating they are dominated by old stellar populations. 
In opposite, the star-forming galaxies are dominated by 
younger stellar populations with \dindex\ less than 1.5. 
This is an expected result since the classifications of star forming, green valley and quiescent galaxies are based on 
sSFR, and there is an anti-correlation between \dindex\ and sSFR \citep{kauffmann-2004}.
This is consistent with the fact that star-forming misaligned 
galaxies are blue disk-like galaxies while green valley and quiescent 
misaligned galaxies are red and bulge-dominated galaxies with lower SFR.

\subsubsection{Radial profiles of \dindex\ \& \hda} \label{sec: radprofile}

In this section, we present the comparison of radial profiles of \dindex\ and \hda\ between kinematically misaligned galaxies and their control samples.

\begin{figure*}
  \begin{center}
    \epsfig{figure=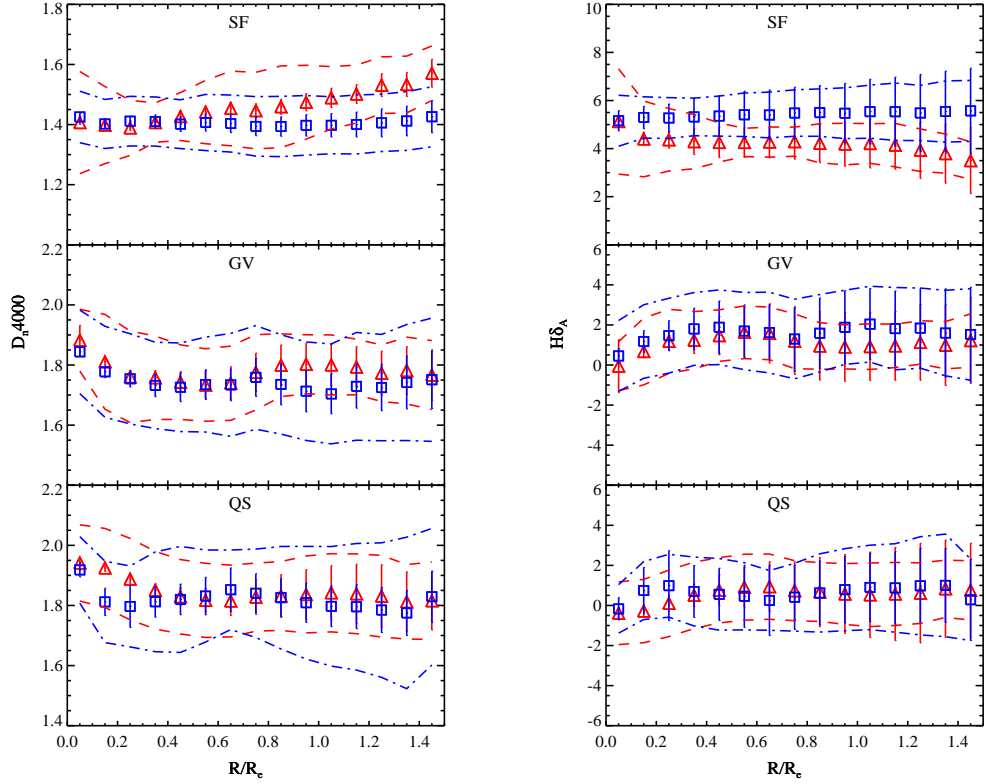,clip=true,width=0.8\textwidth}  
    
  \end{center}
  \caption{Comparison of \dindex\ and \hda\ radial profiles between kinematically misaligned sample (red triangles) and the control sample (blue squares). From top to bottom, three panels show the radial profiles of star-forming (top), green valley (middle) and quiescent (bottom) misaligned galaxies and their control samples. In each panel, the symbols show the mean values of the indices in bins of 0.1$R_e$. The dashed lines and dash-dot lines show the $\pm1\sigma$ scatters for misaligned and control samples, respectively. The error bar shows the typical error of the indices in each bin.}
  \label{fig:radprofile}
\end{figure*}

In Figure~\ref{fig:radprofile}, we show \dindex\ and \hda\ as a function of radius within 1.5 $R_e$ for star-forming, green valley and quiescent misaligned galaxies (red triangles) as well as their control sample (blue squares). The symbols show the mean values in bins of 0.1 $R_e$. The dashed lines and dash-dot lines show the $\pm1\sigma$ scatter region for kinematically misaligned and control samples, respectively. The radii are the projected distances from the galactic center to the spaxels where indices are measured, in the unit of the effective radius.

As the \dindex\ radial profiles show, both the green valley/quiescent misaligned galaxies and their control samples have negative gradients. The stellar populations within 1.5 $R_e$ are old with \dindex\ $\sim$1.8. 
In contrast, the star-forming misaligned galaxies have obvious positive gradient in \dindex, indicating the stellar populations in the central regions are younger than those at outskirts, while the control sample has a flat distribution of \dindex. \cite{chen-prep} introduce the star formation activity parameter to quantify the growth of the misaligned galaxies, finding the rapid growth of the central regions in these galaxies by the acquisition of counter-rotating gas.

\subsection{Gas-phase metallicity}\label{sec: gas-metal}

The chemical abundance is powerful in constraining the evolution and star formation histories of galaxies. Comparing with the stellar metallicity derived from absorption features, the gas-phase metallicity derived from the strong nebular emission lines have the following advantages \citep{Tremonti-04}: (1) the S/N of emission lines exceed that of continuum, and many optically faint galaxies exhibit the highest emission-line equivalent widths. (2) the metallicity measured from emission lines is free of the uncertainty caused by age and $\alpha$-enhancement which plague the interpretation of absorption-line indices. (3) the gas-phase metallicity can reflect the present-day abundance rather than the average of the past generations of stellar populations.
However, the prevalent strong-line abundance diagnostics are developed using the stellar population synthesis and photoionization models, which limits these estimations to be only applicable to the HII regions \citep{kewley-2002}. 

For the first step, we use the BPT diagram \citep{bpt-1981} to separate areas with the AGN excitations, SF excitations, and the combination of them as well as the LINER-like excitations by the criterion described in \cite{kewley-2001} and \cite{kauffmann-2003c}. Only the spaxels with reliable emission line detection (S/N $>$ 3) are used.

For the misaligned star forming galaxies, they experience on-going star formation in the central regions \citep{chen-prep}. Therefore, we directly match them to the MPA/JHU catalog\footnote{http://wwwmpa.mpa-garching.mpg.de/SDSS/DR7/oh.html} to get the metallicity \citep{Tremonti-04} for eight star forming misaligned galaxies.
For green valley/quiescent misaligned galaxies, we find 7 of them have regions dominated by ionization from star-formation, 5 in green valley and 2 in quiescent sequence. Except for two galaxies in which the star formation regions are at the center, all the others have extended star-forming regions ranging from 0.3 to 1.3 $R_e$. 
Because most of the star-forming regions in the 7 green valley/quiescent galaxies are located in the outskirts and the star-forming spaxels distribute close to each other or in a ring, we stack the spectra from star-forming regions in each galaxy to improve S/N, and derive the metallicity from stacked spectra.
Here we use exactly the same metallicity indicator as that used in \cite{Tremonti-04}. 

\begin{figure*}
  \begin{center}
    \epsfig{figure=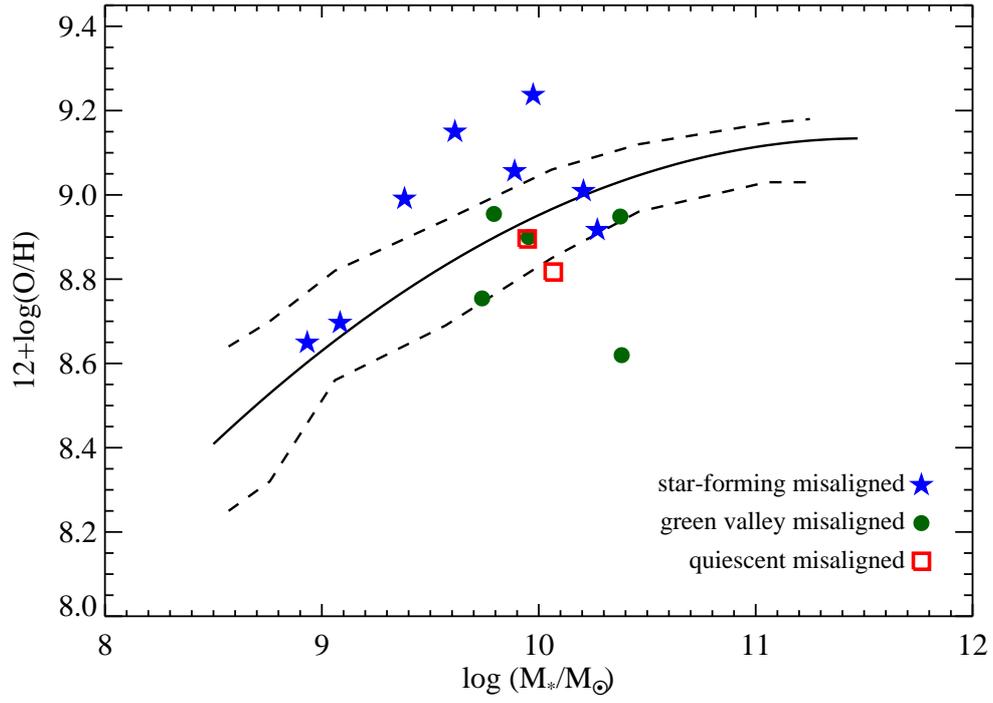,clip=true,width=0.8\textwidth}  
    
  \end{center}
  \caption{The stellar mass-metallicity relation. The solid line show the stellar mass-metallicity relation for the local star forming galaxies given by \protect\cite{Tremonti-04}. The two dashed lines show the $\pm1\sigma$ scatter region. The color symbols show the 8 star-forming (blue star), 5 green valley (green dot) and 2 quiescent (red square) misaligned galaxies.}
  \label{fig:metal}
\end{figure*}

Figure~\ref{fig:metal} shows the stellar mass-metallicity relation. The solid line is the stellar mass$-$metallicity relation for the local star forming galaxies given by \cite{Tremonti-04}, the two dashed lines show the $\pm1\sigma$ scatter region. For the star-forming galaxies (blue stars), we find that half of them follow the typical mass$-$metallicity relation, the other half lie above this relation by 
about 0.3$\sim$0.4 dex. In contrast, the green valley/quiescent galaxies (green dots/red squares) are systematically below the typical mass$-$metallicity relation by about 0.1 dex.

Considering that the spaxels we used in green valley/quiescent galaxies are 0.3$\sim$1.3 $R_e$ away from the center, the universal negative metallicity gradient (0.1 dex per $R_e$) may contribute to the lower metallicity we observed.  
On the other hand, there are only 7 green valley/quiescent galaxies have metallicity measurements and they might not be the representatives for the whole green valley/quiescent misaligned population. Considering the gas in most misaligned green valley/quiescent galaxies is not excited by star formation, we apply an alternative indicator \nii/\sii\ to trace the metallicity. \nii/\sii\ is a good indicator because nitrogen is a secondary $\alpha$-process element while sulphur is the primary nucleosynthesis element \citep{kewley-2002, dopita-2013}. 
Although \nii/\sii\ is the proxy for N/O ratio, in the range of 12+log(O/H) > 8.0, it correlates with O/H abundance very well \citep{dopita-2016, kashino-2016}. Also, the two emission lines are close in wavelength so the reddening is negligible.

\begin{figure*}
  \begin{center}
    \epsfig{figure=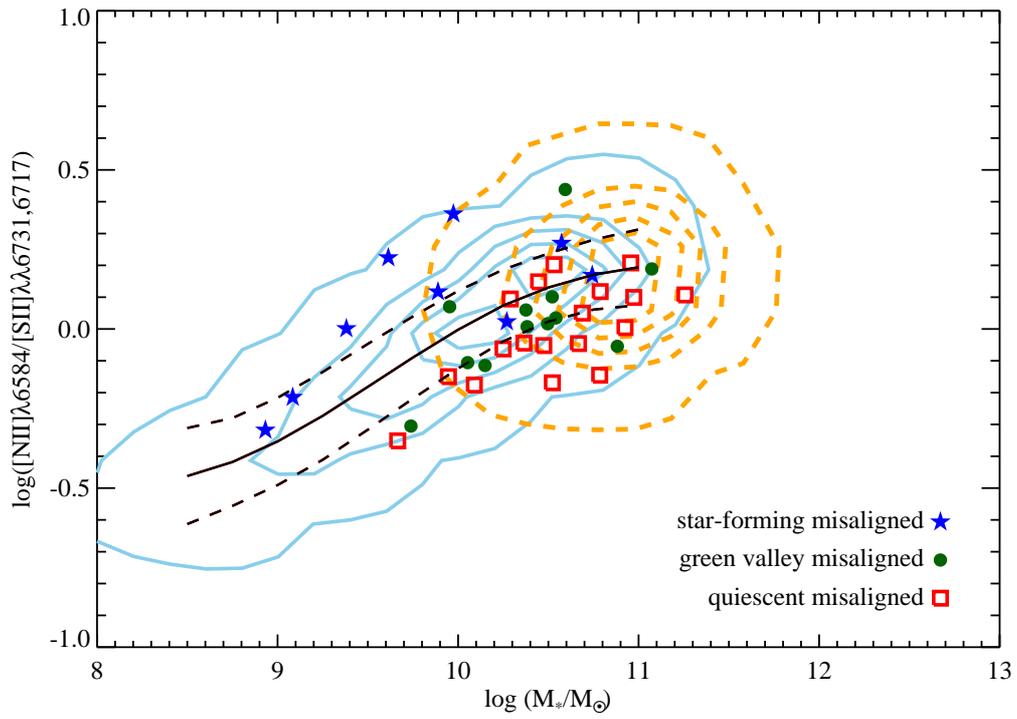,clip=true,width=0.8\textwidth}  
    
  \end{center}
  \caption{The correlation between stellar mass and \nii/\sii. Blue solid and orange dashed contours are for the star-forming galaxies and LINERs in SDSS DR7 database, respectively. Black solid line shows the median while the two dashed lines mark $\pm 1\sigma$ uncertainty region. The color symbols represent different types of kinematically misaligned galaxies.}
  \label{fig:mz}
\end{figure*}

In Figure~\ref{fig:mz}, we show $\log M_*$ vs. $\log$ (\nii/\sii) for SDSS star forming galaxies (blue contour) and LINERs (orange contour) for the central 3 arcsec region. The data are taken from MPA/JHU DR7 catalog. The black solid line shows the median value and two dashed lines show the $\pm1\sigma$ scatter region. There is a strong correlation between stellar mass and \nii/\sii. The change of 0.1 dex in \nii/\sii\ responds to less than an order-of-magnitude change in stellar mass. The LINERs overlap with the star forming galaxies. The over-plotted symbols are the kinematically misaligned galaxies. Consistent with what we see in Figure~\ref{fig:metal}, the star-forming misaligned galaxies (blue stars) have higher \nii/\sii. While misaligned green valley/quiescent galaxies are systematically $\sim$0.1 dex lower than the median value of the whole galaxy sample at fixed stellar mass.

As we know, the gas in the kinematically misaligned galaxies comes from external processes, i.e. gas accretion or mergers. If the gas is accreted from cosmic web, the metallicity should be $\sim$1 dex lower than the typical value. The fact that it is only $\sim$0.1 dex lower in metallicity for green valley and quiescent misaligned galaxies indicates that the gas is accreted from companions which are smaller in mass. This is consistent with \cite{cheung-2016}
, in which they find a kinematically misaligned galaxy is accreting materials from its less massive satellite. 
\cite{davis-2015} show that ETGs with signatures of recent minor mergers have low dust-to-gas ratio as well as metallicity, and typically undergo a recent 10-30:1 merger. This is also consistent with our current result as well as \cite{cheung-2016}. 
However, if the progenitors contain considerable amount of gas, the pre-existing gas would collide with the accreted gas, leading to gas inflow and following star formation in the central regions. The higher central metallicity in star-forming misaligned galaxies could be due to the enrichment from the star formation. In a closed-box model \citep{dalcanton-2007}, the metallicity will mainly depend on the gas mass fraction $f_{\rm gas}\equiv(M_{\rm gas}/(M_{\rm gas} + M_{\rm stars}))$, so the abundances get elevated very quickly as a large fraction of the available gas turns into stars. The low \dindex\ at the center in the star forming misaligned galaxies is a hint that those stars exist.

\subsection{The environment}\label{sec: environment}


In this section, we compare the environment of the misaligned galaxies and the control sample to investigate the environmental dependence of the kinematic misalignment.  

\subsubsection{Estimating the environment}\label{sec:estimate_envir}

We characterize the environment with two parameters, the neighbor number ($N$) and the tidal strength parameter ($Q_{lss}$). The neighbor number is defined as the counts of galaxies brighter than -19.5 mag in $r$-band absolute magnitude 
within a fixed volume of 1 Mpc in projected radius and 500 km s$^{-1}$ in redshift to the primary galaxy. Given the neighbor number is independent of the stellar mass and cannot account for the interaction a galaxy suffering from its satellites, we also use the tidal strength parameter $Q_{lss}$ to depict the effect of total interaction strength produced by all the neighbors within the fixed volume \citep{verley-2007,maria-2015}, the higher the parameter is, the stronger the interaction is. The parameter $Q_{lss}$ is defined as
\begin{equation} \label{equ:qlss}
Q_{lss} \equiv  log\left[\sum_{i} \frac{M_{i}}{M_{p}}\left(\frac{D_{p}}{d_{i}}\right)^{3}\right]
\end{equation}
where $M_{i}$ and $M_{p}$ are the stellar masses of the $i^{th}$ neighbor and the primary galaxy. The $d_{i}$ is the projected distance from the primary galaxy to the $i^{th}$ satellite and $D_{p}$ is the estimated diameter of the central galaxy \citep{maria-2015}. 

For the misaligned sample, about 28.8 per cent (19 out of 66) of them have no neighbor brighter than $-$19.5 mag within the fixed volume (3 star-forming galaxies, 10 green valley galaxies and 6 quiescent galaxies). The relevant numbers for the control samples are 4, 8 and 8, respectively, corresponding to a fraction of $\sim$30 per cent in total.

\begin{figure*}
  \begin{center}
    \epsfig{figure=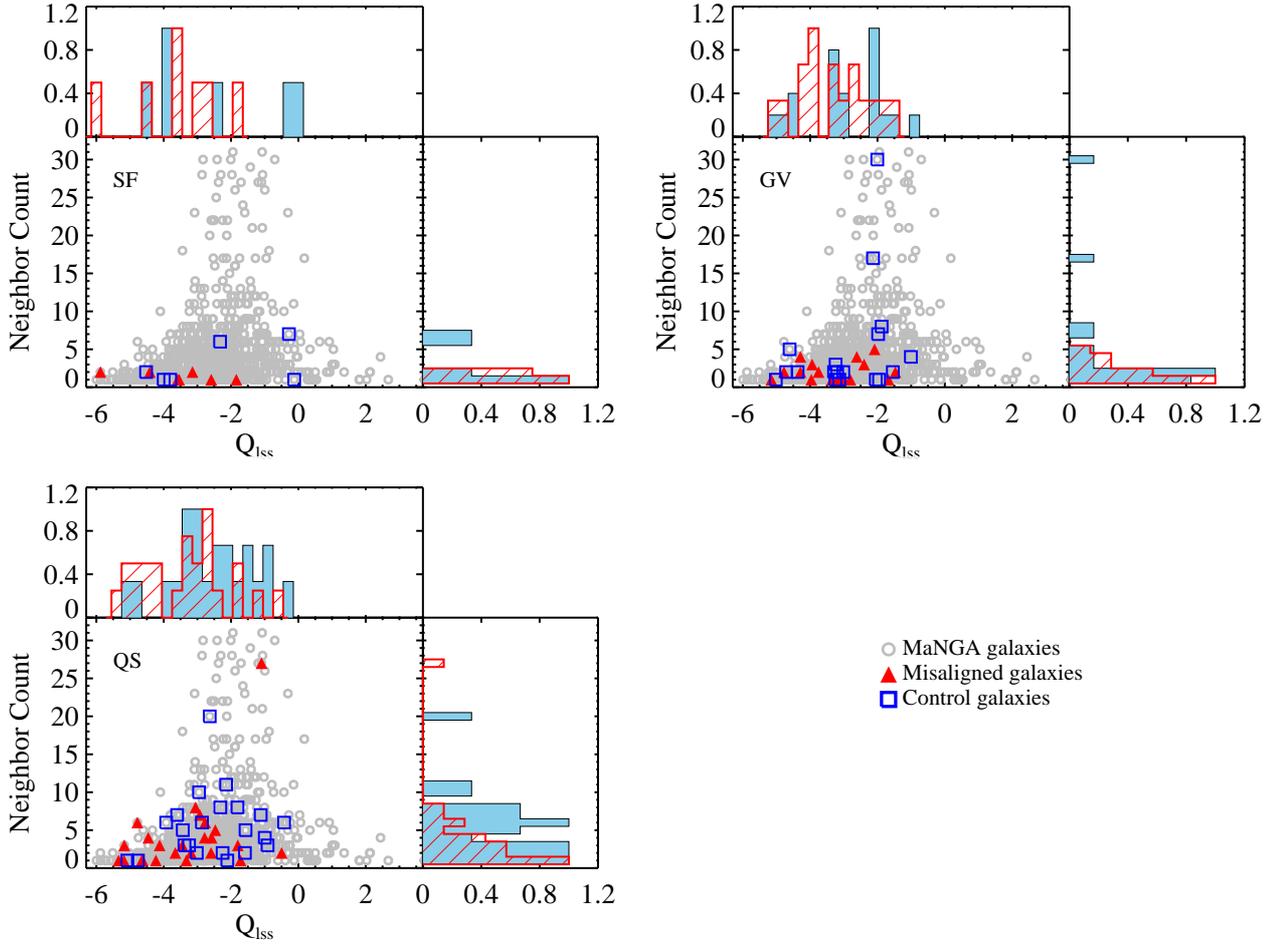,clip=true,width=1\textwidth}  
    
  \end{center}
  \caption{The $Q_{lss}$ parameter versus neighbor counts diagrams. Three panels show the diagrams for star-forming, green valley and quiescent misaligned galaxies, respectively. In each panel, the grey open circles represent 1220 MaNGA galaxies. The kinematically misaligned galaxies and the control sample are presented as red triangles and blue squares, respectively. The top histogram gives $Q_{lss}$ distribution and the right panel shows the distribution of neighbor counts. The histogram filled with red lines is for the kinematically misaligned sample and that filled with blue color is for the control sample. The peaks of all distributions are set to 1.}
  \label{fig:qlss}
\end{figure*}

We present the two environment parameters in  Figure~\ref{fig:qlss}. Again we split the sample into star forming main sequence (top left), green valley (top right) and quiescent sequence (bottom). There is a trend that the kinematically misaligned galaxies (red triangles) are more isolated than the aligned control sample (blue squares) for all the three categories. 
The same trend has been reported by \cite{katkov-2014} where they found around 71 per cent gaseous discs in isolated S0 galaxies are counter-rotating with stellar components.
This is also consistent with the result from ATLAS$^{\rm3D}$ \citep{davis-2011}, in which they find about 42 per cent of fast-rotating field ETGs have kinematically decoupled gas and stars while the fraction is much lower ($\sim$10 per cent) for galaxies within Virgo cluster. 
For a better comparison with the result of \cite{davis-2011}, we match our samples with the group catalog\footnote{http://gax.shao.ac.cn/data/Group.html} generated by \cite{yang-2007}, finding only 1 star-forming and 3 quiescent misaligned galaxies located in clusters (clusters are defined as halo mass $>10^{14}$ \msun). And the fraction of kinematically misaligned galaxies decreases with increasing halo mass.

The above facts are consistent with what \cite{davis-2011} proposed that the dense environment would suppress the external processes of gas acquisition. However, due to the limitation of our sample size, the environmental dependence of misaligned galaxies is not very remarkable. Fortunately, the final MaNGA sample will include $\sim$10,000 galaxies, in which we expect $\sim$500 galaxies to present kinematic misalignment. 

\subsubsection{Galaxy pair}

In this Section, we further investigate whether the kinematic misalignment is caused by close interaction, by quantifying the frequency of close companions. It is well known that the SDSS spectroscopic sample suffers from the incompleteness for close pairs less than 55$^{\prime\prime}$ due to the fiber collision. Fortunately, many SDSS regions are observed by multiple plates such that some close pairs can be recovered. It has been shown that with proper correction with spectroscopic completeness, it is possible to conduct a statistical study of close pairs with the SDSS sample \citep{patton-2008}. In what follows, we ignore the spectroscopic incompleteness effect, since we are only interested in comparing the pair fraction between the kinematically misaligned sample and its corresponding control sample in a relative sense. 

For each galaxy, we search for its spectroscopic companion(s) with projected separation $<$ 50 kpc/h and line-of-sight velocity difference $<$ 500 km s$^{-1}$. 
We find 10 ($\sim15 per cent$) misaligned galaxies are in pairs, 2 star-forming, 4 green valley as well as 4 quiescent, and the number of corresponding control sample in pairs are 2, 1 and 4, respectively. 
It seems that the kinematic misalignment is not necessarily associated with galaxy pairs, except for the misaligned green valley galaxies.
Again, we caution that the sample size is quite limited. A larger sample is required to draw a more robust conclusion.

\section{Discussion}\label{sec:discussion}


The properties and origins of kinematically misaligned galaxies have been studied from the 
perspective of observations \citep{kannappan-2001, Sarzi06, davis-2011} and simulations \citep{voort-2015, lagos-2015}. 
However, previous studies are mainly focusing on ETGs (ellipticals and S0) but rarely in star-forming galaxies. 

Thanks to the large complete MaNGA sample, \cite{chen-prep} for the first time detect kinematic misalignment in star-forming galaxies 
in a large complete sample. 
The fraction of misaligned galaxies detected in star-forming main sequence is $\sim$2 per cent, while the fraction in green valley and quiescent sequence is $\sim$7 per cent and $\sim$8 per cent , respectively.

We investigate the morphology, stellar population, metallicity as well as environment of different types of misaligned galaxies 
and find the star-forming misaligned galaxies have different properties from the green valley/quiescent ones.

\subsection{Properties of kinematically misaligned galaxies}

The morphologies of different types of misaligned galaxies are separated clearly by s$\acute{\rm e}$rsic index. 
About 90 per cent misaligned ones in green valley and quiescent sequence are early type galaxies with $\emph{n}>2$. 
In contrast, the misaligned star-forming galaxies are more disky, having lower s{\'e}rsic index. 
\cite{chen-prep} fit the $r$-band surface brightness profiles of the misaligned star-forming galaxies and find half of them 
already have photometric bulge components. 


We use \dindex\ and \hda\ to trace the local stellar populations.
The green valley and quiescent misaligned galaxies are dominated by old stellar populations with \dindex\ $\sim$1.8 and they have similar negative radial gradients as the control sample. Given that wet major merger is an efficient way in triggering star formation \citep{veilleux-2002,springel-2005a,kormendy-1992}, the high \dindex\ value suggests no wet major mergers occurred in green valley/quiescent misaligned galaxies within the last 1$\sim$2 Gyr. The star-forming misaligned galaxy sample has positive \dindex\ gradient with young stellar populations with \dindex\ $<$1.5 at the central regions, while the control sample has a flat distribution of \dindex\ on average.

We use both 12+Log(O/H) and \nii/\sii\ as gas-phase metallicity tracers, finding the quiescent and green valley misaligned galaxies are $\sim$0.1 dex poorer in metal than the prediction of typical stellar mass-metallicity relation derived from local star-forming galaxies. But the value is not as low as that of the pristine gas from cosmic web. We propose the kinematically decoupled gas in green valley/quiescent galaxies is accreted from satellites which are smaller in stellar mass. 
\cite{cheung-2016} studied one misaligned galaxy in detail, finding it is stripping material from a companion. The situation is opposite in star-forming misaligned galaxies, where the metallicity is higher than the typical stellar mass-metallicity relation. Due to the intense star-forming activities at the central regions, large fraction of gas turns into stars and as a consequence elevates the abundance instantaneously.

By comparing the environments of the misaligned galaxies with a control sample closely matched in SFR $\&$ \mstar\ , we find that the misaligned galaxies prefer to locate at lower density areas. In quality, this result is consistent with that of \cite{davis-2011}, in which they find that the galaxies in Virgo cluster are mostly kinematically aligned between gas and stars while around 42$\pm$5 per cent field fast-rotating ETGs host kinematically misaligned gas. They propose the dense environment may suppress the host galaxies accreting external kinematically decoupled gas. 

\subsection{Origins of kinematically misaligned galaxies}\label{sec:discuss_origin}

The origin of kinematically misaligned galaxies is a puzzle for a long time. 
Given the angular momentum conservation, it's natural to suppose external 
processes to play the key role in leading kinematic misalignment between gas and stars. 
Simulations show that episodic and continuous gas accretion 
\citep{thakar-1996, thakar-1998} as well as merging with companions 
\citep{bois-2011, naab-2014} can lead to kinematically decoupled components in the host galaxies. 

For the misaligned galaxies located in the star forming main sequence, most of them have positive \dindex\ gradients. 
\cite{chen-prep} propose a scenario to explain this phenomenon. 
The gas-rich progenitors accrete external kinematically decoupled gas. 
The inevitable gas-gas collisions between accreted and pre-existing gas lead to the cancellation of angular momentum, accelerating 
the gas inflow. If the angular momentum carried by the external gas exceeds that 
in pre-existing gas, the kinematic misalignment would show up in the host galaxies. On the other hand, 
large amounts of gas sinks to the central region, leading to the fast centrally-concentrated star formation 
and instantaneous elevation of metallicity. This picture explains why the misaligned galaxies located in the star 
forming main sequence have positive radial profiles in \dindex\ and higher metallicity at galaxy centers.
The stars formed through this process will finally become bulge components although at the beginning 
they belong to a disk from a kinematic point of view. In other words, this process 
leads to the fast growth of central parts of galaxies as \cite{chen-prep} suggested.

Most misaligned galaxies in green valley/quiescent sequence have old stellar population with \dindex\ $\sim$1.8. 
Both the misaligned galaxies and control sample have similar negative radial 
gradients in \dindex. And the metallicities in these galaxies are $\sim$0.1 dex lower 
than the prediction of local stellar mass-metallicity relation. 
In these green valley or quiescent galaxies, the progenitors might be passive gas-poor ETGs. 
They accrete misaligned gas from neighbors that are smaller in mass. 
On one hand, the amount of the accreted gas can only rejuvenate ETGs with low level star-formation activity.
On the other hand, the small collision cross section between gas and stars leads to a long dynamical friction timescale.
The gas-star misalignment will persist for 2-5 Gyr \citep{voort-2015, davis-2016} in these systems leading to a higher fraction of misaligned galaxies.


As we mentioned above, the accretion of large amount of misaligned gas is an efficient way for the growth of central part of galaxies (final bulge). 
Here comes the question $-$ whether bulges formed from 
misaligned gas acquisition have similar properties as those formed from the ordinary way, 
such as major merger. This question is out of the scope of the current paper, but should be an interesting 
project in near future.

\section{Summary}\label{sec:summary}

We study 66 kinematically misaligned galaxies selected from 1351 MaNGA survey.
These galaxies are classified into three categories according to 
their sSFR, 10 of them are star-forming galaxies, 26 located in the green valley and 30 are 
quiescent galaxies. The kinematically decoupled galaxies located in the star forming main sequence 
appear to have different properties, i.e. stellar populations, metallicity, from green valley and quiescent ones.

\hangafter =1 

\hangindent = 1em
\noindent$\bullet$ We study how the fraction of kinematically misaligned galaxies varies with galaxy 
physical parameters, i.e. \mstar, SFR, sSFR, finding the fraction peaks at log\mstar/\msun $\sim$ 10.5, and it decreases with SFR $\&$ sSFR.

\hangindent = 1em
\noindent$\bullet$ From the perspective of morphology, kinematic 
decoupled galaxies located in green valley and quiescent sequence appear to be bulge dominated. 
The star-forming misaligned galaxies are more disky. 
For all the kinematically misaligned galaxies, their ionized gaseous and stellar velocity fields are regular. 
However, their \ewhae\ maps can be classified into four different morphological types, i.e. peaks at center, 
biconical patterns, ring-like structures and no obvious structures.

\hangindent = 1em
\noindent$\bullet$ The stellar populations traced by \dindex\ and \hda\ show that stars in 
quiescent and green valley kinematically misaligned galaxies are old. The values of \dindex\ and \hda\ 
indicate that these galaxies haven't experienced substantial star formation in the past 1$\sim$2 Gyr. 
The kinematically misaligned galaxies located in the star forming main sequence have younger 
stellar populations and fast star formation at the central regions.

\hangindent = 1em
\noindent$\bullet$ In quiescent and green valley kinematically misaligned galaxies, the gas-phase 
metallicity is $\sim$0.1 dex lower than the prediction of local stellar mass$-$metallicity relation. While in star forming 
misaligned galaxies, half of them have 0.3$-$0.4 dex higher central gas-phase metallicity.

\hangindent = 1em
\noindent$\bullet$ All types of kinematically misaligned galaxies prefer to 
locate in low density areas, indicating that dense environment would suppress the external process of gas acquisition 
and the formation of misaligned galaxies.

In the kinematically misaligned star-forming galaxies, the 
accreted gas collides with the pre-existing gas, losing angular momentum and triggering star formation in 
central regions, leading to the fast growth of the central regions. 
In contrast, the green valley/quiescent misaligned 
galaxies are formed in the way that gas-poor progenitors accrete kinematically decoupled gas from dwarf satellites. 

\section*{Acknowledgements}

Y.M.C acknowledge support from NSFC grant 11573013, 11133001, the Opening Project of Key Laboratory of Computational Astrophysics, National Astronomical Observatories, Chinese Academy of Sciences. Y.S. acknowledges support from NSFC grant 11373021, the CAS Pilot-b grant No. XDB09000000 and Jiangsu Scientific Committee grant BK20150014. C.A.T. acknowledges support from National Science Foundation of the United States Grant No. 1412287.

\noindent This work was supported by World Premier International Research Center Initiative (WPI Initiative), MEXT, Japan and by JSPS KAKENHI Grant Number JP15K17603

\noindent Funding for the Sloan Digital Sky Survey IV has been provided by the Alfred P. Sloan Foundation, the U.S. Department of Energy Office of Science, and the Participating Institutions. SDSS- IV acknowledges support and resources from the Center for High-Performance Computing at the University of Utah. The SDSS web site is www.sdss.org.

\noindent SDSS-IV is managed by the Astrophysical Research Consortium for the Participating Institutions of the SDSS Collaboration including the Brazilian Participation Group, the Carnegie Institution for Science, Carnegie Mellon University, the Chilean Participation Group, the French Participation Group, Harvard-Smithsonian Center for Astrophysics, Instituto de Astrof\'{i}sica de Canarias, The Johns Hopkins University, Kavli Institute for the Physics and Mathematics of the Universe (IPMU) / University of Tokyo, Lawrence Berkeley National Laboratory, Leibniz Institut f\"{u}r  Astrophysik Potsdam (AIP), Max-Planck-Institut f\"{u}r  Astronomie (MPIA Heidelberg), Max-Planck-Institut f\"{u}r Astrophysik (MPA Garching), Max-Planck-Institut f\"{u}r  Extraterrestrische Physik (MPE), National Astronomical Observatory of China, New Mexico State University, New York University, University of Notre Dame, Observat\'{o}rio Nacional / MCTI, The Ohio State University, Pennsylvania State University, Shanghai Astronomical Observatory, United Kingdom Participation Group, Universidad Nacional Aut\'{o}noma de M\'{e}xico, University of Arizona, University of Colorado Boulder, University of Oxford, University of Portsmouth, University of Utah, University of Virginia, University of Washington, University of Wisconsin, Vanderbilt University, and Yale University.







\bsp	
\label{lastpage}
\end{document}